\documentclass[aps,pra,amsfonts,amsmath,amssymb,showpacs,reprint]{revtex4-1}
\usepackage{graphicx, mathtools, hyperref, comment, bm}
\usepackage[caption=false]{subfig}

\newcommand*{\avg}[1]{\langle #1 \rangle} 
\newcommand*{\abs}[1]{\left| #1 \right|} 
\newcommand{\ket}[1]{\left| #1 \right>} 
\newcommand{\braket}[2]{\left< #1 \vphantom{#2} | #2 \vphantom{#1} \right>} 
\newcommand*{\pd}[2]{\frac{\partial #1}{\partial #2}} 
\renewcommand*{\d}[2]{\frac{\mathrm{d} #1}{\mathrm{d} #2}} 
\newcommand{\var}[1]{\mathrm{var} #1}

\begin{document}

\title{Quantum chaos in a Bose-Hubbard dimer with modulated tunnelling}

\author{R. A. Kidd}
\author{M. K. Olsen}
\author{J. F. Corney}
\affiliation{School of Mathematics and Physics, University of Queensland, Brisbane, Queensland 4072, Australia.}

\date{\today}

\begin{abstract}
In the large-$N$, classical limit, the Bose-Hubbard dimer undergoes a transition to chaos when its tunnelling rate is modulated in time. 
We use exact and approximate numerical simulations to determine the features of the dynamically evolving state that are correlated with the presence of chaos in the classical limit. We propose the statistical distance between initially similar number distributions as a reliable measure to distinguish regular from chaotic behaviour in the quantum dynamics. Besides being experimentally accessible, number distributions can be efficiently reconstructed numerically from binned phase-space trajectories in a truncated Wigner approximation. Although the evolving Wigner function becomes very irregular in the chaotic regions, the truncated Wigner method is nevertheless able to capture accurately the beyond mean-field dynamics.

\end{abstract}



\maketitle

\section{\label{sec:intro}Introduction}
Chaos is characterised by an exponential sensitivity to small perturbations in initial conditions (hyperbolicity), non-integrable dynamics, and a widely-traversed phase space (ergodicity)~\cite{Wimberger2014}. Despite its ubiquity in the classical macroscopic world, chaos is not intrinsic to the underlying quantum description, except where an effective nonlinearity is introduced through, e.g., taking a classical limit or mean-field approximation,  or conditioning the dynamics upon the results of a measurement~\cite{Eastman2017}. Nevertheless, the behaviour of quantum states in classically chaotic regimes, referred to as `quantum chaos', can mimic features of classical chaos, at least for short times. Quantum chaos is thus of intrinsic interest for the insight it brings to the quantum-classical crossover~\cite{Haake2010, Gutzwiller1992}, in addition to its connections with ergodicity and thermalisation \cite{Altland2012}.



The sensitivity to initial conditions  in classical chaos is quantified by Lyapunov exponents: a positive Lyapunov exponent indicates an exponential growth in the separation in phase space between trajectories that are initially close together.
A direct implementation of this measure in quantum systems is problematic due to the intrinsic quantum uncertainty: it is not possible deterministically to specify arbitrarily precise initial conditions in phase space. A  naive alternative would be the overlap of initially similar quantum states; however this overlap is time-invariant under unitary evolution. For this reason, the overlap of initially identical distributions but under slightly perturbed Hamiltonians has been used to distinguish between quantum chaos and regular dynamics~\cite{Haake2010}. States with large support on regular regions of phase space will retain close to unitary overlap under a small Hamiltonian perturbation, whereas states that are well-confined to chaotic regions will exhibit an exponential decay of overlap to around order $1/N$~\cite{Haake2010}, where $N$ is the number of particles. This decay may be used to extract effective Lyapunov exponents~\cite{Cucchietti2002}.

However, quantum-state overlap is not directly accessible experimentally. Even theoretically, the quantum state becomes intractable to calculate or even specify exactly for moderately sized systems, due to the exponential scaling of Hilbert-space dimension on the number of modes.

In this paper, we propose and benchmark the use of the overlap of number distributions, rather than the quantum state itself, as a measure of quantum chaos. Number distributions are experimentally accessible in ultra-cold atom experiments, and can be calculated without knowing the full quantum state, such as through the binning of trajectories in a truncated Wigner simulation~\cite{Lewis-Swan2016}. In particular, we calculate the Bhattacharyya distance~\cite{Bhatta1943} between number distributions in a driven, double-well Bose-Einstein condensate, bench-marking approximate approaches against exact calculations. We find that the Bhattacharyya distance -- even when calculated approximately using binned truncated Wigner trajectories -- is a reliable indicator of quantum chaos that is consistent with other measures. The advantage of this method is that it scales well with system size and so may be applied to a wide range of systems for which exact calculations are intractable. Our results also provide further insight into the nature and scope of quantum chaos in a simple model that can be implemented and probed experimentally~\cite{Tomkovic2017}.

We note that an alternative solution is possible for continuously monitored quantum systems~\cite{Eastman2017, Pokharel2018, Eastman2019}, where individual stochastic quantum trajectories, representing the dynamics conditioned on measurement, can be used to derive effective quantum Lyapunov exponents. In this work, we study exclusively the isolated dynamics of the two-site lattice, where such an approach based on continuous measurement readout is not available.

Distributions over phase space provide a powerful way to connect the evolving quantum state to the dynamical structure of the corresponding classical system. For example, the Husimi $Q$-function, besides providing a visualisation of the state~\cite{Mahmud2005}, is the basis for an approximate treatment of the quantum dynamics through an ensemble of classical trajectories~\cite{Trimborn2009, Hennig2012}. Our use of a Wigner function aims to improve on this approach, since to obtain classical Liouvillian evolution, the truncation of terms in the evolution equation is less severe in the Wigner case than in the $Q$-function case.
Although the truncated Wigner method has been used extensively in ultracold atoms, it has not, to our knowledge, been used to study chaotic dynamics in quantum systems.


We begin, in section~\ref{sec:system} by introducing the double-well system and discussing the chaotic dynamics that arises in a semiclassical analysis when the tunnelling rate is modulated in time. In section~\ref{sec:quantum} we consider the quantum dynamics, calculated exactly and with the truncated Wigner method. Finally in section~\ref{sec:DB} we introduce the Bhattacharyya distance as a measure of chaos in quantum systems and test its performance when used with the truncated Wigner representation.

\section{\label{sec:system}Modulated Bose-Hubbard dimer}

For a sufficiently tight trapping potential, a Bose-Einstein condensate in a double-well potential~\cite{Milburn1997} may be modelled as a two-site Bose-Hubbard model, or quantum dimer:
\begin{equation}
    \hat{H} = \hbar U \left( \hat{b}_1^\dagger \hat{b}_1^\dagger \hat{b}_1 \hat{b}_1 + \hat{b}_2^\dagger \hat{b}_2^\dagger \hat{b}_2 \hat{b}_2 \right) - \hbar J \left( \hat{b}_1^\dagger \hat{b}_2 + \hat{b}_2^\dagger \hat{b}_1 \right),
  \label{eq:Hdimer}
\end{equation}
where $J$ is the tunnel-coupling rate and $U$ is the strength of the on-site interaction term, which arises from interparticle collisions treated in the low-temperature $s$-wave limit. The canonical boson operators obey the commutation relations $[\hat b_j, \hat b_k^\dagger] = \delta_{jk}$. To introduce chaos into this system, we will vary the tunnelling rate in time: $J(t) = J_0 + \mu\cos{(\omega t)}$, which could be achieved, for example, through modulation of the barrier height~\cite{Milburn1997chaos}.

Even for thousands of particles, the relevant Hilbert space of the system is small enough to allow for direct calculation of the dynamical wavefunction, affording the opportunity to benchmark approximate approaches against exact solutions; yet the interplay between coherent tunnelling and particle interactions leads to rich many-body physics. This physics is reflected in the nonlinear behaviour of the corresponding semiclassical model, which can undergo macroscopic self-trapping and, for appropriate modulation of parameters, chaotic dynamics.

The semiclassical, or mean-field, model describes the dynamics of the order parameter under the assumption that two-body correlations factorise, $\avg{AB} = \avg{A}\avg{B}$. It is justifiable in the large-particle limit where fluctuations around the mean value of observables are small. It is convenient to define `Bloch-sphere' observables:
\begin{align}
    x &= \avg{\hat{J}_x} =  \frac{\avg{\hat{b}_2^\dagger \hat{b}_1} + \avg{\hat{b}_1^\dagger \hat{b}_2}}{2}, \nonumber \\
    y &= \avg{\hat{J}_y} = \frac{\avg{\hat{b}_2^\dagger \hat{b}_1} - \avg{\hat{b}_1^\dagger \hat{b}_2}}{2i}, \nonumber \\
    z &= \avg{\hat{J}_z} = \frac{\avg{\hat{b}_2^\dagger \hat{b}_2} - \avg{\hat{b}_1^\dagger \hat{b}_1}}{2},
\end{align}
which, in the semiclassical limit, satisfy $x^2+y^2+z^2 = N^2/4$.
The dynamics on the Bloch sphere can be mapped to a plane
$(z,\phi)$, where $z$ is the population difference between the two sites and $\phi = -\arg{(x + iy)}$
is the relative phase. The semiclassical dynamics is then given by the coupled equations:
\begin{align}
    \d{z}{t} &= 2J(t)\sqrt{N^2/4 - z^2} \sin{(\phi)}, \nonumber \\
    \d{\phi}{t} &= -2z \left( \frac{J(t) \cos{(\phi)}}{\sqrt{N^2/4-z^2}} + 2U \right).
    \label{eq:dimerSCEOM}
\end{align}

Without modulation of the parameters, this system is integrable, with solutions given by Jacobi elliptic functions~\cite{Milburn1997}. The nonlinear dynamics is governed by the ratio of interaction energy to the tunnelling rate, $C = UN/J_0$. As $C$ is increased through the critical value $C = 1$, the stable centre at $(z,\phi) = (0,0)$ undergoes a pitchfork bifurcation. The separatrix that emerges from the resultant saddle point encloses each of the two new stable centres and prevents trajectories in one enclosed region from traversing to the other. This nonlinear suppression of tunnelling, known as macroscopic self-trapping, has been observed experimentally~\cite{Albiez2005}.

\begin{figure}[ht]
    \centering
    \includegraphics[width=\columnwidth]{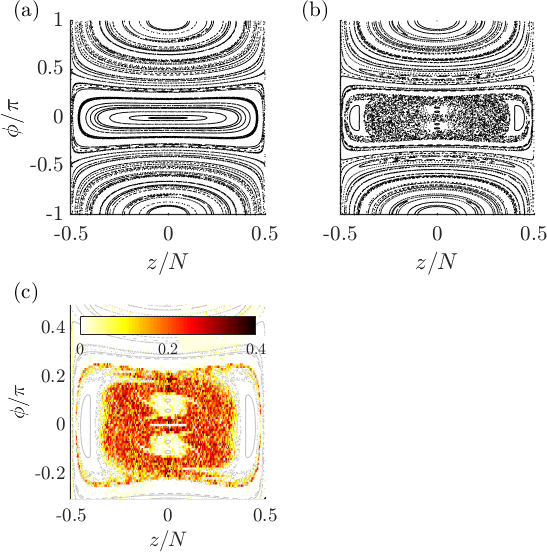}
    \caption{
    Semiclassical dynamics. Stroboscopic Poincar\'e sections for (a) no driving and (b) driving with amplitude $\mu = 0.2$ and frequency $\omega = 1.37J_0$. Points are plotted at intervals of the driving period, $T=2\pi/(1.37J_0)$. The colour-scale plot (c) indicates the finite-time Lyapunov exponents calculated at each point in a truncated phase-space region for the same driving amplitude and frequency as (b). For all plots, and throughout the paper, $C=1$, corresponding to the point of bifurcation.}
    \label{fig:phasespace}
\end{figure}

Modulating the tunnelling rate breaks integrability and gives rise to the possibility of chaos. For example, at the critical value $C = 1$, a chaotic sea emerges in the region of the bifurcating point (see Fig.~\ref{fig:phasespace}) and continues to persist at $C>1$ around the separatrix~\cite{Milburn1997chaos, Mourik2018}. We note that integrability may be broken in other ways, such as modulation of the interaction strength~\cite{Gong2009, Watanabe2012} or the relative height of the wells~\cite{Morales-Molina2008}.

Chaos in phase space can be illustrated by stroboscopic Poincar\'e sections, which are constructed by evolving and periodically plotting a selection of semiclassical trajectories at intervals of the driving period~\cite{Haake2010}. Poincar\'e sections at the bifurcation point are shown in Fig.~\ref{fig:phasespace}a and  Fig.~\ref{fig:phasespace}b, without and with driving, respectively. In Fig.~\ref{fig:phasespace}b a chaotic sea emerges around the bifurcating fixed point, within which sit several Kolmogorov-Arnold-Moser (KAM) islands~\cite{Haake2010}.

The maximal Lyapunov exponent at phase-space coordinate $\bm{z}(0)$ is defined as
\begin{equation}
    \lambda = \lim_{t\to\infty} \lim_{\bm{z}_0' \to \bm{z}_0}  \frac{1}{t} \ln{\left( \frac{\abs{\abs{\bm{z}(t) - \bm{z}'(t)}}} {\abs{\abs{\bm{z}(0) - \bm{z}'(0)}}} \right)},
\end{equation}
where the initial values of the two phase-space trajectories $\bm{z}(0)$ and $\bm{z}'(0)$ should in principle be made vanishingly close to each other. In practice, we implement this numerically by choosing a small but finite difference and simulate for a time that is long enough for the maximal exponent to dominate the rate of divergence (but small enough that this divergence is not limited by phase-space boundaries)~\cite{Chianca2011, Wimberger2014}. Fig.~\ref{fig:phasespace}c was generated by evolving pairs of perturbed points distributed across phase space and performing a linear fit on their logarithmic relative divergence, $\ln{\left( \frac{\abs{\abs{\bm{z}(t) - \bm{z}'(t)}}} {\abs{\abs{\bm{z}_0 - \bm{z}'_0}}} \right)}$ against time, $t$. The initial perturbation magnitude was $\abs{z - z'} = 10^{-4}$ and the trajectories were evolved for 20 driving periods. The evolution time of the finite-time Lyapunov exponents was small enough to avoid saturation of the logarithmic relative divergence. The darker regions of Fig.~\ref{fig:phasespace}c, which represent large positive Lyapunov exponents, correspond well with the chaotic regions of phase space shown in the Poincar\'e section, Fig.~\ref{fig:phasespace}b. The white regions in Fig.~\ref{fig:phasespace}c correspond to non-positive Lyapunov exponents.

\begin{figure}[ht]
    \centering
    \includegraphics[width=\columnwidth]{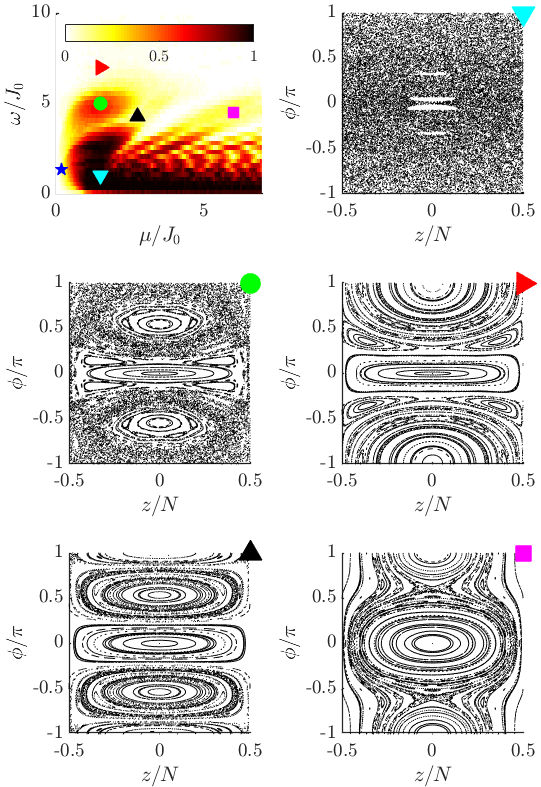}
    \caption{Top left: the fraction of the phase space occupied by the chaotic sea as a function of driving amplitude $\mu$ and frequency $\omega$. The remaining plots are Poincare sections for the various choices of modulation parameters $(\mu,\omega)$ indicated by the symbols. Extensive chaos is observed in Fig.~\ref{fig:chaos_regimes}$\blacktriangledown$, which gradually disappears with increasing modulation frequency through Fig.~\ref{fig:chaos_regimes}$\bullet$ and Fig.~\ref{fig:chaos_regimes}$\blacktriangleright$. Three roughly equal-sized stable islands appear in Fig.~\ref{fig:chaos_regimes}$\blacktriangle$, with minimal chaos about the separatrices. The star $\bigstar$ indicates the parameters used in Fig.~\ref{fig:phasespace}b.}
    \label{fig:chaos_regimes}
\end{figure}

The extent of chaos in the modulated dimer can be controlled via the driving amplitude, $\mu$, and frequency, $\omega$. Fig.~\ref{fig:chaos_regimes} quantifies the fraction of phase space occupied by chaotic regions. For this plot, we calculated the finite-time Lyapunov exponents for 1600 trajectories distributed across the phase space and classified each phase-space point as chaotic if its Lyapunov exponent was greater than the largest Lyapunov exponent within the unmodulated phase space. The percentage of phase space that was chaotic was then determined and plotted for varying modulation amplitude and frequency. We note that this approach does not produce exact Lyapunov exponents, yet is sufficient to distinguish regular and chaotic trajectories to a high degree of accuracy in all tested parameter regimes.

Even though the modulation is a requirement for chaotic behaviour in this system, driving at high frequency tends to diminish the extent of the chaos. We illustrate this Fig.~\ref{fig:chaos_regimes} by plotting several representative Poincar\'e sections highlighting the different regimes. For example, a sufficiently rapid modulation ($\omega \gtrsim 6 J_0$) will suppress chaos entirely, whereas for low modulation frequency ($\omega \sim 0.5 J_0$), widespread chaos persists for modulation amplitudes as high as $\mu = 100 J_0$.


\begin{figure}[ht]
    \centering
    \includegraphics[width=\columnwidth]{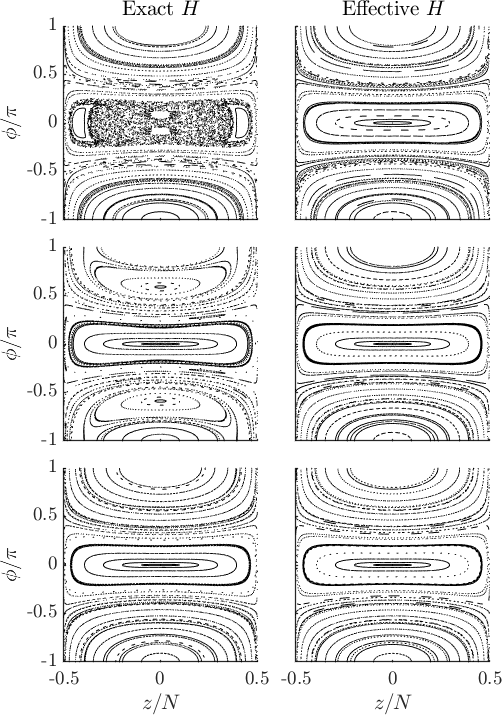}
    \caption{Poincar\'e sections corresponding to (left) exact Hamiltonian~\eqref{eq:Hdimer} and (right) high-frequency effective Hamiltonian~\eqref{eq:H_highf}. Modulation amplitude is $\mu = 0.2 J_0$ for all plots and the rows from the top correspond to modulation frequencies of $\omega/J_0 = 1.37, 5, 10$.}
    \label{fig:high_frequency}
\end{figure}

\begin{figure}[ht]
    \centering
    \includegraphics[width=\columnwidth]{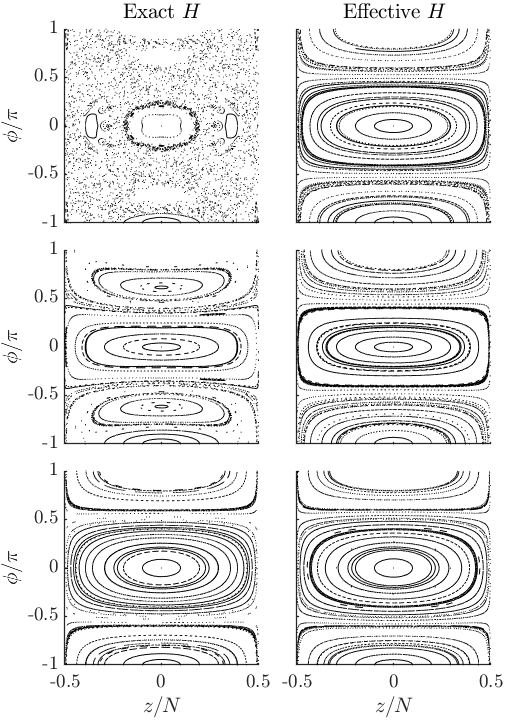}
    \caption{Poincar\'e sections corresponding to (left) exact Hamiltonian~\eqref{eq:Hdimer} and (right) high-frequency effective Hamiltonian~\eqref{eq:H_highf}. Modulation amplitude is $\mu = 9.57925 J_0$; all other parameters as in Fig.~\ref{fig:high_frequency}}
    \label{fig:high_frequency2}
\end{figure}

Consistent with the disappearance of chaos at high modulation frequency observed in Fig.~\ref{fig:chaos_regimes}, an effective static integrable Hamiltonian~\cite{Kohler2005, Gong2009} can be derived under the assumption that the modulation frequency is much larger than other frequency scales in the system:
\begin{align}
    \hat{H}_\text{eff} &= \frac{\hbar U}{4} \left[3+\mathcal{J}_0 \left(\frac{4\mu}{\omega} \right)\right] \left( \hat{b}_1^\dagger \hat{b}_1^\dagger  \hat{b}_1 \hat{b}_1 + \hat{b}_2^\dagger \hat{b}_2^\dagger \hat{b}_2 \hat{b}_2 \right) \nonumber\\
    &\quad - \frac{\hbar U}{4} \left[1-\mathcal{J}_0 \left(\frac{4\mu}{\omega} \right)\right] \Big( \hat{b}_1^\dagger \hat{b}_1^\dagger  \hat{b}_2 \hat{b}_2 + \hat{b}_2^\dagger \hat{b}_2^\dagger \hat{b}_1 \hat{b}_1 \nonumber\\
    &\qquad\qquad - 4 \hat{b}_1^\dagger \hat{b}_1 \hat{b}_2^\dagger \hat{b}_2 \Big)
    -\hbar J_0\left( \hat{b}_1^\dagger \hat{b}_2 + \hat{b}_2^\dagger \hat{b}_1 \right)
    \label{eq:H_highf}
\end{align}
where $\mathcal{J}_0$ is the zeroth order Bessel function of the first kind (calculations in App.~\ref{sec:Floquet_work}). As shown in Figs.~\ref{fig:high_frequency} and \ref{fig:high_frequency2}, the Poincar\'e sections corresponding to the time-periodic and effective Hamiltonians become very similar for sufficiently large modulation frequency ($\omega = 10$).

\section{\label{sec:quantum}Quantum dynamics}


For the two-mode system, the semiclassical results above can be compared to the exact quantum dynamics, which can be calculated through a Floquet analysis. Such an approach was used, for example, to investigate dynamical tunnelling between regular islands in the semiclassical Poincare section~\cite{Salmond2002, Watanabe2012}. 

\begin{figure}[ht]
    \centering
    \includegraphics[width=\columnwidth]{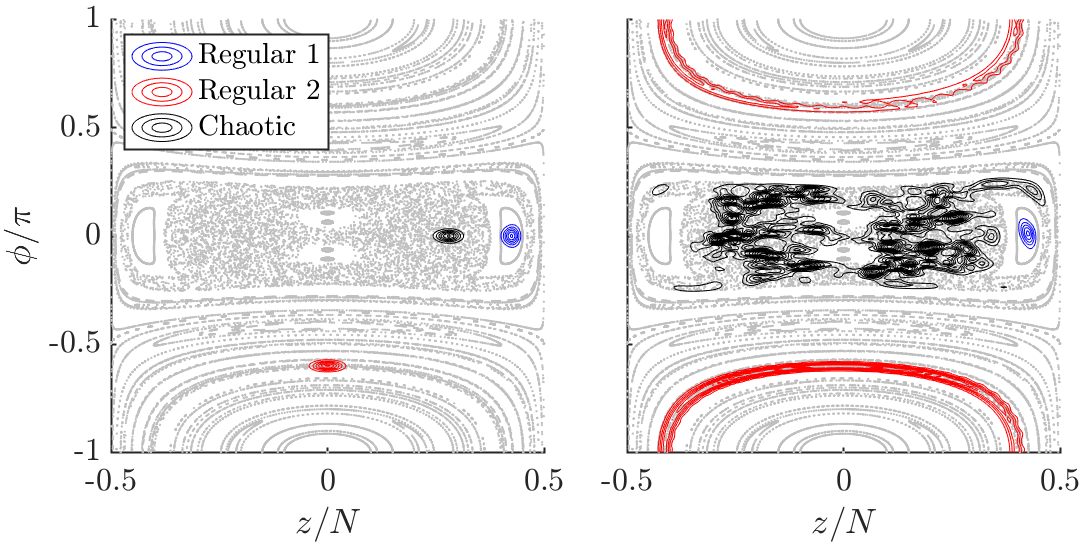}
    \caption{$Q$ functions for (a) initial coherent states and (b) after 20 driving periods overlaid on semiclassical Poincar\'e section. The total particle number is $N=1000$ and the other parameters are as in Fig.~\ref{fig:phasespace}b.}
    \label{fig:Q_t}
\end{figure}

To make a connection with the classical Poincare sections, we represent the evolving quantum state in phase space through use of the atomic $Q$ function~\cite{Arecchi1972, Lee1993}. Fig.~\ref{fig:Q_t} shows the $Q$ function of what are initially Bloch coherent states \footnote{Bloch coherent states~\cite{Arecchi1972} are minimal uncertainty states of the dimer under the restriction of fixed total particle number. They may be created experimentally by, for example, forming a condensate of known number in one well, followed by linear tunnelling~\cite{Tomkovic2017}.} centred at three representative points in phase space.


After 20 driving periods, the $Q$ function initially centred in the chaotic region has become very irregular and is smeared out over nearly all of the chaotic sea, reflecting the ergodic nature of the corresponding classical trajectories~\cite{Mahmud2005}. By contrast, the $Q$ function of the `regular 1' state remains localised on the KAM island on which it was initially placed. The `regular 2' state is smeared out, but in a qualitatively very different way from that seen in the chaotic region; the dependence of the tunnelling oscillation period upon its amplitude causes the distribution to be sheared along the regular classical trajectories (a two-mode analogue of the optical Kerr effect).


\begin{figure}[ht]
    \centering
    \includegraphics[width=0.95\columnwidth]{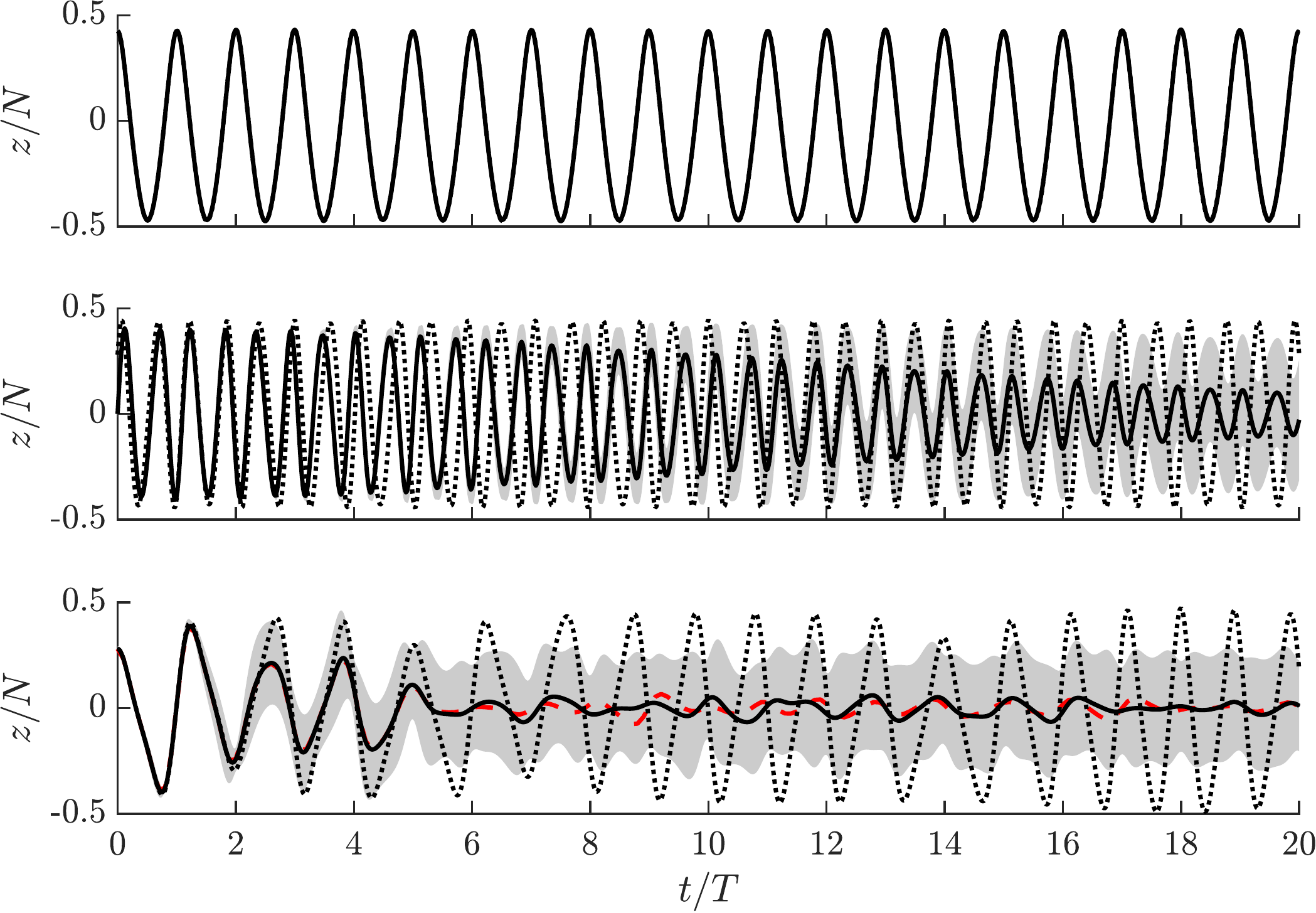}
    \caption{Population difference $z$ as a function of time for the regular 1 (top), regular 2 (middle), and chaotic states (bottom), calculated exactly (solid lines), with truncated Wigner method (dashed red lines) and semiclassically (dotted lines). The exact standard deviation of the population difference is indicated by grey shading. The ensemble standard error of all truncated Wigner results, which used 10,000 paths, is less than the line width.}
    \label{fig:z_t}
\end{figure}

When the $Q$ function is no longer well-localised, it is clear that the semiclassical predictions -- which neglect fluctuations about the mean -- will become unreliable. This neglect of fluctuations in the chaotic regime is rapidly manifest in discrepancies in the predictions for observables, as illustrated in Fig.~\ref{fig:z_t}. In the chaotic regime, the semiclassical simulation shows persistent, albeit irregular, tunnelling oscillations in the number difference $\avg{\hat{z}}$ whereas in the quantum dynamics this oscillation rapidly decays. The decay in the mean of $z$ is associated with a rise in the variance, which supports the interpretation of the chaotic quantum state as an ensemble of irregular classical trajectories that get out of phase over time. On a much longer time scale, the tunnelling oscillations of the `regular 2' state also collapse, but this is due to the gradual dephasing of regular oscillations at slightly different frequencies.


The fact that a fairly rapid growth of variance occurs in the `regular 2' as well as the chaotic regions indicates that the variance itself is not always a reliable probe of the presence of chaos~\cite{Tomkovic2017}. Nevertheless, when growth in variance is driven by chaotic behaviour, it does reveal something about the extent of the chaotic region.


For the two-mode system, the spreading of the initially coherent $Q$ function over the Bloch sphere is directly linked with the fragmentation of the condensate~\cite{Trimborn2009}. Fragmentation occurs when not all particles occupy the same single particle state -- as required for a pure BEC -- and is quantified by the eigenvalues of the one-particle reduced density matrix~\cite{Leggett2001},
\begin{equation}
    \hat{\rho}_\text{red} = \begin{pmatrix} \avg{\hat{b}_1^\dagger\hat{b}_1} & \avg{\hat{b}_1^\dagger\hat{b}_2} \\ \avg{\hat{b}_2^\dagger\hat{b}_1} & \avg{\hat{b}_2^\dagger\hat{b}_2} \end{pmatrix}.
\end{equation}
A fragmented condensate is revealed when $\hat{\rho}_\text{red}$ has more than one nonzero eigenvalue, i.e., when $\lambda_\text{max} <N$.

For the two-mode system, the maximum eigenvalue of $\hat{\rho}_\text{red}$ can be written in terms of the length of the Bloch spin vector $|\avg{\hat{\bf{J}}}|$~\cite{Trimborn2009}, which in turn can be expressed in terms of the variance of the Bloch variables:
\begin{align}
    \lambda_\text{max} &= \frac{N}{2} + |\avg{\hat{\bf{J}}}| \nonumber \\
    |\avg{\hat{\bf{J}}}| &= \sqrt{\frac{N(N+2)}{4} - \var{(\hat{J}_x)} - \var{(\hat{J}_y)} - \var{(\hat{J}_z})}.
    \label{eq:bloch_spin}
\end{align}
Since fragmentation is linked to the growth in the variance of any of the Bloch variables, it provides a  more global picture of the spread of the state over the Bloch sphere than the variance in any one particular variable.


\begin{figure}[ht]
    \centering
    \includegraphics[width=\columnwidth]{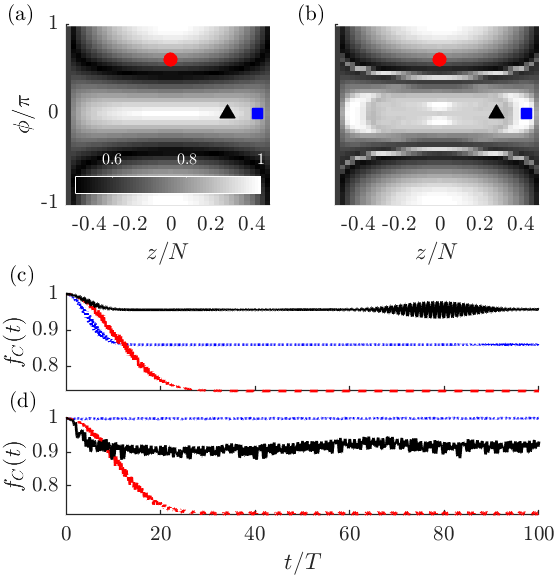}
    \caption{Condensate fraction after evolution of initial Bloch states without modulation (panels (a) and (c)), and with modulation $\mu = 0.2 J_0$  (panels (b) and (d)). Upper panels give the condenste fraction map in phase space after a time of $20T$. Lower panels show the time evolution of the condensate fraction for the three representative states:  regular 1 (blue dotted lines / $\blacksquare$), regular 2 (red dashed lines / $\bullet$) and chaotic (black solid lines / $\blacktriangle$)}
    \label{fig:condensate_fraction}
\end{figure}

The largest eigenvalue is plotted as a function time in Fig.~\ref{fig:condensate_fraction} for the three representative initial conditions, both with and without modulation. The first thing to note is that loss of coherent tunnelling oscillations does not mean total fragmentation: the maximum eigenvalue for the state in the chaotic sea falls only to about $0.9N$, far above the lower limit of $0.5N$. Second, fragmentation is not just a consequence of chaotic behaviour. Indeed, the largest fragmentation occurs in the regular region where the classical dynamics exhibit large-amplitude tunnelling oscillations~\cite{Hennig2012} and for which the corresponding $Q$ functions are sheared across a large solid angle on the Bloch sphere. The $Q$ function of the state in the chaotic region -- although it spreads out to subtend a similarly large angle on the Bloch sphere -- actually corresponds to less fragmentation, since its $Q$ function fills in the intermediate region of phase space and thus maintains a higher level of total average spin. For larger relative nonlinearity $C$ or larger modulation amplitude $\mu$, where the chaotic sea occupies a greater fraction of phase space~\cite{Trimborn2009} (see Fig.~\ref{fig:chaos_regimes}), we see increased fragmentation from the chaotic dynamics.

Third, perhaps the most striking consequence of the modulation is the near total suppression of fragmentation in the $Q$ function of the `regular 1' state. This suppression of fragmentation is due to the localisation enforced by creation of the KAM islands, and is particularly clear in  Fig.~\ref{fig:condensate_fraction}b, in which the regions of high purity (low fragmentation) match closely the centres of the KAM islands seen in in Fig.~\ref{fig:phasespace}.

Overall then, in regions of phase space dominated by chaotic behaviour, the structures in the Poincare section determine the level of fragmentation in the quantum dynamics. Conversely, if the system is known to be chaotic, then fragmentation in the quantum system could be used to determine the presence of KAM islands or to probe the size of the chaotic sea, even though fragmentation by itself is not an indicator of the presence of chaos.


By choosing system parameters to precisely alter the size of the chaotic sea, it is possible to generate states with arbitrary condensate fraction from initial coherent states on the timescale of few modulation periods.

\subsection{\label{sec:TW}Truncated Wigner method}


Thus far we have calculated the quantum dynamics through exact calculation of the state vector, with a subsequent mapping to phase space via the $Q$ function. We now use these results to benchmark the truncated Wigner method as a tool for investigating quantum chaos. Not only is the truncated Wigner method an efficient way to simulate many-mode quantum systems (the exact state-vector approach quickly becomes intractable even for just a few modes), it gives a direct access to dynamics in phase space, in which much of the classical analysis of chaotic systems is framed.

In the truncated Wigner method~\cite{Graham1973}, a Fokker-Planck equation for an approximate Wigner function is obtained by truncation of the full partial differential equation for the Wigner function at second-order. This approximation is justified by the relatively small size of the coefficients of the third-order terms for a system with many particles. The Fokker-Planck equation is then mapped through standard techniques~\cite{Gardiner1985} onto a set of ordinary differential equations for stochastic complex amplitudes, $\beta_j$. Although the Wigner function is not positive-definite for all states, a positive approximation to common initial states~\cite{Olsen2009} allows the low-order moments to be accurately sampled. Any such approximation to the initial state together with the truncation of the third-order terms neglects any negativity of the Wigner function, but gives accurate operator moments up to the collapse timescale of quantum optics~\cite{Corney2015}.




The truncated Wigner equations for the Bose-Hubbard dimer are:
\begin{align}
  \d{\beta_1}{t} &= iJ \beta_2 - 2iU \left( \abs{\beta_1}^2 - 1 \right) \beta_1, \nonumber \\
  \d{\beta_2}{t} &= iJ \beta_1 - 2iU \left( \abs{\beta_2}^2 - 1 \right) \beta_2,
  \label{eq:dimerTwEOM}
\end{align}
which
are solved for an ensemble of trajectories. Expectation values of symmetrically ordered products can be calculated directly as stochastic averages, for example,
\begin{equation}
  \frac{1}{2} \avg{\hat{b}_i^\dagger \hat{b}_j + \hat{b}_j \hat{b}_i^\dagger} = \int\! \beta_i^* \beta_j \, W(\vec\beta^*, \vec\beta) \, \mathrm{d}^{4}\beta
  .
\end{equation}

The truncated Wigner simulations used in this work were implemented with the XMDS2~\cite{Dennis2013} software package.

The truncated Wigner calculation of the expectation value of the two-mode number difference, $z$, is included in Fig.~\ref{fig:z_t}. The results for the two states in regular regions are indistinguishable from the exact calculations over this time scale. Some discrepancy is seen for the state in the chaotic region, but only well after the collapse of the tunnelling oscillations.

\begin{figure}[ht]
    \centering
    \includegraphics[width=\columnwidth]{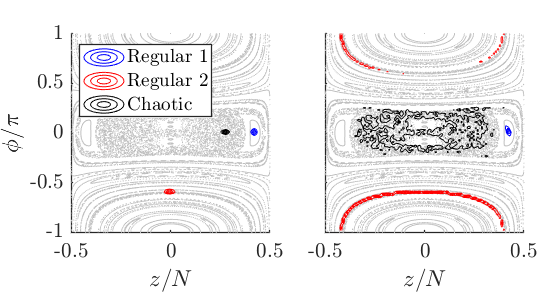}
    \caption{Initial coherent state reconstructed $W$-functions ($10^4$ truncated Wigner trajectories) before and after 20 driving periods overlaid on the Poincar\'e section. Note that the reconstructed $W$-function, unlike the true $W$-function, is all-positive.}
    \label{fig:W_t}
\end{figure}

Binning trajectories allows us to build up a picture of the phase-space distribution without having to calculate the full density matrix. For large $N$, the exact Wigner function on the Bloch sphere is particularly difficult to calculate, as it requires generation of high-order spherical harmonics~\cite{Trimborn2009}. Fig.~\ref{fig:W_t} shows the binned distribution after 20 driving periods for the three representative states, showing a close correspondence to the $Q$ functions calculated in Fig.~\ref{fig:Q_t}.



\section{\label{sec:DB}Bhattacharyya distance}

Sensitivity to initial conditions is the hallmark of chaos in classical systems. Although a direct measure of the divergence of particular phase-space trajectories is not accessible in unitary quantum dynamics, we have seen above that the divergence is reflected in the way that $Q$ and Wigner functions within the chaotic region spread out to cover the whole of the chaotic sea. Moreover, they evolve in a very irregular way that contrasts markedly with the regular shearing, for example, that occurs for the `regular 2' state.

\begin{figure}[ht]
  \centering
  \includegraphics[width=\columnwidth]{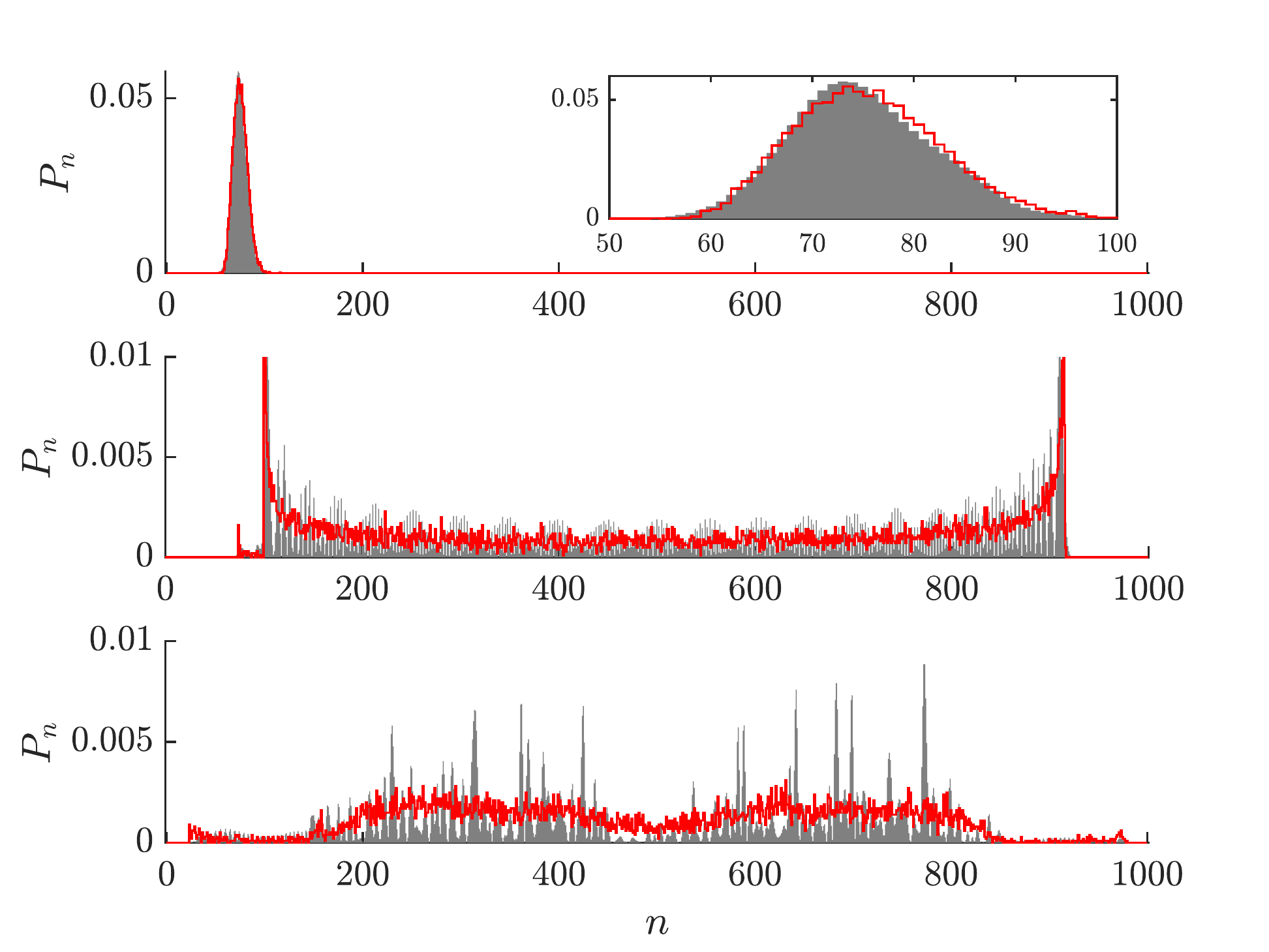}
  \caption{Number distribution for site 1 for the regular 1 (upper), regular 2 (middle) and chaotic (lower) states in Fig.~\ref{fig:Q_t} after 20 driving periods. The grey histogram is the exact number distribution and the red line is an approximate number distribution constructed from binning $10^4$ truncated Wigner trajectories.}
  \label{fig:number_dist}
\end{figure}

To obtain a quantitative measure of the difference between the irregular and regular evolution, we calculate the overlap of number distributions. The irregular evolution of the number distribution in the chaotic region (see the lower panel of Fig.~\ref{fig:number_dist}) can be expected to lead to a loss of overlap between number distributions that are initially similar, or between number distributions that are initially identical but which evolve under slightly different Hamiltonians.

We consider the overlap of number distributions, rather than the overlap of the entire state~\cite{Grass2013}, since (a) number distributions are much more experimentally accessible, and (b) in a theoretical calculation it may be feasible to calculate the number distribution even when the full state is intractable. In particular, we benchmark the use of binned truncated Wigner trajectories~\cite{Lewis-Swan2016} to calculate the number distribution approximately, since this method remains tractable for large numbers of particles and modes.

For two states expressed in the number-state basis, $\ket{\psi} = \sum_n c_n \ket{n}$ and $\ket{\psi'} = \sum_n d_n \ket{n}$, the state overlap is
\begin{equation}
  O(\psi,\psi') = \abs{\braket{\psi}{\psi'}} = \abs{\sum_{n,m} c^*_n d_m \braket{n}{m}} = \abs{\sum_{n} c^*_n d_n},
\end{equation}
whereas the overlap of number distributions $P_n = \abs{c_n}^2 $ and $\tilde P_n = \abs{d_n}^2 $ is given by
\begin{equation}
  B(\psi,\psi') = \sum_n \sqrt{P_n \tilde{P_n}} = \sum_{n} \abs{c^*_n d_n}.
\end{equation}
The triangle inequality holds that $B(\psi,\psi') \geq O(\psi,\psi')$. 
Although the number distribution overlap is not equivalent to state overlap, it provides an upper bound and can function as a practical alternative.

To obtain a quantum analogue of the maximal Lyaponuv exponent, we calculate the Bhattacharyya statistical distance~\cite{Bhatta1943}:
\begin{equation}
  D_B(\psi,\psi') = -\ln{\left[B(\psi,\psi')\right]},
  \label{eq:DB}
\end{equation}
developed as a means of characterising the divergence between two probability distributions and which
takes the minimum value of zero when the number distributions are identical.


Following Ref.~\cite{Lewis-Swan2016}, we construct the approximate number distribution for site 1 by calculating the proportion $P_n$ of truncated Wigner trajectories that fall into bins satisfying
\begin{equation}
    n  \leq \left|\beta_1\right|^2 < n + 1.
\end{equation}

We can expect this binning approach to be accurate when the Wigner function does not change rapidly over the bin size and for a large number of particles such that the corrections from symmetric ordering can be neglected.
%
%
Fig.~\ref{fig:number_dist} gives a comparison of the exact and reconstructed number distributions for the three representative states after 20 driving periods.
In all cases, the binned reconstructions recover the overall extent and the coarse features of the distributions. On the other hand, the fine features of the regular 2 and chaotic states are lost in the reconstructions, which raises the question of how sensitive to chaos the Bhattacharyya distance based on these distributions will be.



\begin{figure}[ht]
    \centering
    \includegraphics[width=0.95\columnwidth]{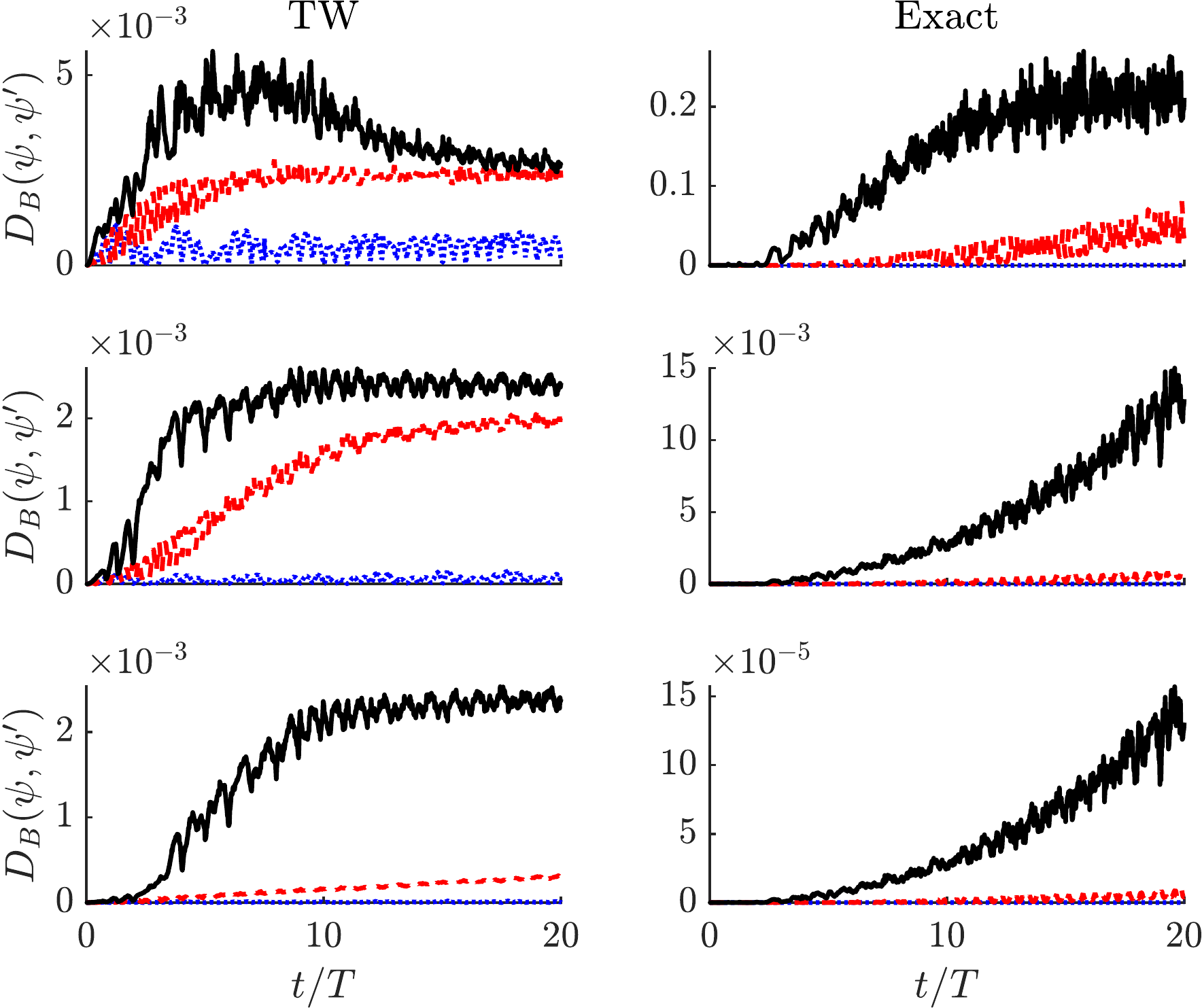}
    \caption{Bhattacharyya distances between number distributions due perturbation of the interaction term, $U' = (1+p)U$, calculated from (left) $10^5$ binned truncated Wigner trajectories and (right) exact state vector. Initial states are as in Fig.~\ref{fig:Q_t}: `regular 1' (blue dotted line), `regular 2' (red dashed line), `chaotic' (black solid line). The perturbation amplitude for each row is, from the top, $p = 10^{-3}, 10^{-4}, 10^{-5}$.}
    \label{fig:DB_p}
\end{figure}

We calculate the Bhattacharyya distance between distributions evolved under Hamiltonians that differ slightly in the strength of the interaction: $U$ and $U' = (1+p)U$, for $p\ll1$. Fig.~\ref{fig:DB_p} shows the results for the binned trajectories calculation (left) and the the direct state-vector calculation (right). Despite the smoothing effect seen in Fig.~\ref{fig:number_dist}, the Bhattacharyya distance for the chaotic state is distinctly larger than that of the regular states, so long as a sufficiently small perturbation is chosen. However, the Bhattacharyya distances determined via the truncated Wigner and exact state methods quantitatively differ, likely due to the loss of fine features in the number distribution evident in the truncated Wigner reconstruction shown in Fig.~\ref{fig:number_dist}.


Moreover, the truncated Wigner Bhattacharyya distance for the chaotic state saturates to about $D_B \approx 2.4\times10^{-3}$ for perturbations of $p=10^{-4},10^{-5}$, behaviour which is not evident for the corresponding exact state Bhattacharyya distances. This saturation is consistent with the statistical uncertainty in the binning process. For an estimate of the saturation limit, consider a uniform number distribution $P_n = \frac{1}{N+1}$ and a perturbed distribution $\tilde{P}_n = P_n + \epsilon$, where $\epsilon$ is a Gaussian random variable with mean zero and standard deviation $\sigma$; the Bhattacharyya distance has analytic value $D_B = -\ln{\left(1 - \frac{(N+1)^2}{8} \sigma^2\right)}$ in the limit of large $N$. For $D_B \approx 2.4\times10^{-3}$, this estimate gives $\sigma \approx 1.4 \times 10^{-4}$, which is approximately consistent with a Poissonian process sorting $N_t = 10^5$ trajectories into $N_b = 10^3$ bins with mean bin occupation $\bar{p} = 10^2$ and occupation standard deviation $\sigma_p = 10$ (Poissonian):
\begin{align}
    \frac{\sigma_p}{\bar{p}} &= 10^{-1}, &
    \frac{\sigma}{\bar{P}_n} &\approx 1.4\times10^{-1}.
\end{align}
Choosing $10^4$ trajectories causes the Bhattacharyya distance to saturate at around $D_B \approx 2.4\times10^{-2}$, also consistent with a Poissonian process for this reduced number of trajectories. Therefore, we can conclude that the saturation of the Bhattacharyya distance is due to sampling error in the binning process.

\section{\label{sec:conc}Conclusions}


In summary, we have explored the quantum and semiclassical dynamics of the modulated Bose-Hubbard dimer around the point at which the semiclassical model undergoes a bifurcation. In the semiclassical model, the extent of the chaos can be controlled by variation of both the modulation frequency and amplitude. Chaotic behaviour tends to be suppressed at high modulation frequency, consistent with the effective static integrable Hamiltonian that can be derived in this regime.

The diverging semiclassical trajectories in the chaotic region correspond, in the quantum dynamics, to phase-space distributions that spread out over time to fill the entire chaotic sea. For a two-mode system, a broad phase-space distribution on the Bloch sphere, corresponding to a diminished average spin vector, correlates directly with fragmentation. We found, nevertheless, that it was states in the regular region undergoing nonlinear shearing that had the greatest loss of condensate fraction (i.e. greater fragmentation).

Despite the complex behaviour of the underlying phase-space distributions, we found that an ensemble of truncated Wigner trajectories reproduced the behaviour of exact quantum expectation values of chaotic quantum states to a high degree of accuracy.

To obtain a measure for quantum systems that plays a similar role that Lyaponuv exponents do in classical chaos, we considered the divergence of number distributions caused by perturbations to the Hamiltonian, as quantified by the Bhattacharyya distance. We benchmarked the use of binned truncated Wigner trajectories as an approximation to the number distribution that is extensible to large-$N$, many site lattices, where exactly calculating the state overlap becomes intractable. We found that for sufficiently small perturbations to the Hamiltonian, the Bhattacharyya distance reliably distinguished between quantum states localised in the regular region from those in the chaotic region. This is in spite of the lack of quantitative agreement in the values of the Bhattacharyya distance calculated with binned Wigner trajectories compared with those calculated exactly. We attribute this discrepancy to the truncated Wigner approach smoothing over some of the fine features of the true number distribution.


\appendix

\section{\label{sec:Floquet_work} Effective Hamiltonian for high-frequency modulation}

To derive the effective Hamiltonian~\eqref{eq:H_highf}, we introduce the pseudospin operators $\hat{J}_{x,y,z}$ satisfying commutation relations $[\hat{J}_j, \hat{J}_k] = i \epsilon_{jkl} \hat{J}_l$, where $j,k,l\in\{x,y,z\}$ and $\epsilon_{jkl}$ is the Levi-Civita symbol,
\begin{align}
    \hat{J}_x &= \frac{\hat{a}_2^\dagger \hat{a}_1 + \hat{a}_1^\dagger \hat{a}_2}{2}, &
    \hat{J}_y &= \frac{\hat{a}_2^\dagger \hat{a}_1 - \hat{a}_1^\dagger \hat{a}_2}{2i}, \nonumber\\
    \hat{J}_z &= \frac{\hat{a}_2^\dagger \hat{a}_2 - \hat{a}_1^\dagger \hat{a}_1}{2}, &
    \hat{N} &= \hat{a}_1^\dagger \hat{a}_1 + \hat{a}_2^\dagger \hat{a}_2.
\end{align}
The time-independent $H_0$ and time-dependent $H_1$ parts of the Hamiltonian~\eqref{eq:Hdimer} can then be written as:
\begin{eqnarray}
    \hat{H}_0 &=& 2U \hat{J}_z^2 - 2J_0\hat{J}_x + \frac{U}{2} \hat{N} \left( \hat{N} - 2 \right)\nonumber \\
    \hat H_1 &=& 2\mu\cos{(\omega t)} \hat{J}_x.
\end{eqnarray}

The effective Hamiltonian~\cite{Kohler2005, Gong2009} under a high-frequency modulation of period $T = 2\pi/\omega$ is:
\begin{equation}
    \hat{H}_\text{eff} = \frac{1}{T} \int_0^T e^{iA(t) \hat{J}_x} \hat{H}_0 e^{-iA(t) \hat{J}_x} \, \mathrm{d}t,
    \label{eq:effH}
\end{equation}
where we have written $\int_0^t \! \hat{H}_1(t')\,\mathrm{d}t' = A(t) \hat{J}_x$, with
$A(t) \equiv \frac{2\mu}{\omega} \sin{(\omega t)}$.

To evaluate the self-interaction term in Eq.~(\ref{eq:effH}), we can use the Baker-Hasdorff lemma~\cite{Sakurai2014}:
\begin{align}
    e^{iA(t) \hat{J}_x}& \hat{J}_z^2 e^{-iA(t) \hat{J}_x} \nonumber\\
    &= \frac{\cos{(2A(t))}}{2} \left( \hat{J}_z^2 - \hat{J}_y^2 \right) \nonumber\\
    &\quad + \frac{\sin{(2A(t))}}{2}  \left( \hat{J}_y\hat{J}_z + \hat{J}_z\hat{J}_y \right) \nonumber\\
    &\quad + \frac{1}{2} \left( \hat{J}_z^2 + \hat{J}_y^2 \right)
\end{align}
together with the integrals:
\begin{align}
    \int_0^T \cos{(2A(t))} \, \mathrm{d}t &= 2\pi \mathcal{J}_0\left(\frac{4\mu}{\omega}\right), \\
    \int_0^T \sin{(2A(t))} \, \mathrm{d}t &= 0,
\end{align}
where $\mathcal{J}_0(z)$ is the zeroth order Bessel function of the first kind.


\begin{thebibliography}{35}%
\makeatletter
\providecommand \@ifxundefined [1]{%
 \@ifx{#1\undefined}
}%
\providecommand \@ifnum [1]{%
 \ifnum #1\expandafter \@firstoftwo
 \else \expandafter \@secondoftwo
 \fi
}%
\providecommand \@ifx [1]{%
 \ifx #1\expandafter \@firstoftwo
 \else \expandafter \@secondoftwo
 \fi
}%
\providecommand \natexlab [1]{#1}%
\providecommand \enquote  [1]{``#1''}%
\providecommand \bibnamefont  [1]{#1}%
\providecommand \bibfnamefont [1]{#1}%
\providecommand \citenamefont [1]{#1}%
\providecommand \href@noop [0]{\@secondoftwo}%
\providecommand \href [0]{\begingroup \@sanitize@url \@href}%
\providecommand \@href[1]{\@@startlink{#1}\@@href}%
\providecommand \@@href[1]{\endgroup#1\@@endlink}%
\providecommand \@sanitize@url [0]{\catcode `\\12\catcode `\$12\catcode
  `\&12\catcode `\#12\catcode `\^12\catcode `\_12\catcode `\%12\relax}%
\providecommand \@@startlink[1]{}%
\providecommand \@@endlink[0]{}%
\providecommand \url  [0]{\begingroup\@sanitize@url \@url }%
\providecommand \@url [1]{\endgroup\@href {#1}{\urlprefix }}%
\providecommand \urlprefix  [0]{URL }%
\providecommand \Eprint [0]{\href }%
\providecommand \doibase [0]{http://dx.doi.org/}%
\providecommand \selectlanguage [0]{\@gobble}%
\providecommand \bibinfo  [0]{\@secondoftwo}%
\providecommand \bibfield  [0]{\@secondoftwo}%
\providecommand \translation [1]{[#1]}%
\providecommand \BibitemOpen [0]{}%
\providecommand \bibitemStop [0]{}%
\providecommand \bibitemNoStop [0]{.\EOS\space}%
\providecommand \EOS [0]{\spacefactor3000\relax}%
\providecommand \BibitemShut  [1]{\csname bibitem#1\endcsname}%
\let\auto@bib@innerbib\@empty
\bibitem [{\citenamefont {Wimberger}(2014)}]{Wimberger2014}%
  \BibitemOpen
  \bibfield  {author} {\bibinfo {author} {\bibfnamefont {S.}~\bibnamefont
  {Wimberger}},\ }\href {\doibase 10.1007/978-3-319-06343-0_1} {\emph {\bibinfo
  {title} {Nonlinear Dynamics and Quantum Chaos}}},\ Graduate Texts in Physics\
  (\bibinfo  {publisher} {Springer International Publishing},\ \bibinfo {year}
  {2014})\BibitemShut {NoStop}%
\bibitem [{\citenamefont {Eastman}\ \emph {et~al.}(2017)\citenamefont
  {Eastman}, \citenamefont {Hope},\ and\ \citenamefont
  {Carvalho}}]{Eastman2017}%
  \BibitemOpen
  \bibfield  {author} {\bibinfo {author} {\bibfnamefont {J.~K.}\ \bibnamefont
  {Eastman}}, \bibinfo {author} {\bibfnamefont {J.~J.}\ \bibnamefont {Hope}}, \
  and\ \bibinfo {author} {\bibfnamefont {A.~R.~R.}\ \bibnamefont {Carvalho}},\
  }\href {\doibase 10.1038/srep44684} {\bibfield  {journal} {\bibinfo
  {journal} {Sci. Rep.}\ }\textbf {\bibinfo {volume} {7}},\ \bibinfo {pages}
  {44684} (\bibinfo {year} {2017})}\BibitemShut {NoStop}%
\bibitem [{\citenamefont {Haake}(2010)}]{Haake2010}%
  \BibitemOpen
  \bibfield  {author} {\bibinfo {author} {\bibfnamefont {F.}~\bibnamefont
  {Haake}},\ }\href {\doibase 10.1007/978-3-642-05428-0} {\emph {\bibinfo
  {title} {Quantum Signatures of Chaos}}},\ \bibinfo {edition} {3rd}\ ed.,\
  \bibinfo {series} {Springer Series in Synergetics}, Vol.~\bibinfo {volume}
  {54}\ (\bibinfo  {publisher} {Springer-Verlag Berlin Heidelberg},\ \bibinfo
  {year} {2010})\BibitemShut {NoStop}%
\bibitem [{\citenamefont {Gutzwiller}(1992)}]{Gutzwiller1992}%
  \BibitemOpen
  \bibfield  {author} {\bibinfo {author} {\bibfnamefont {M.~C.}\ \bibnamefont
  {Gutzwiller}},\ }\href {\doibase 10.1038/scientificamerican0192-78}
  {\bibfield  {journal} {\bibinfo  {journal} {Sci. Am.}\ }\textbf {\bibinfo
  {volume} {266}},\ \bibinfo {pages} {78} (\bibinfo {year} {1992})}\BibitemShut
  {NoStop}%
\bibitem [{\citenamefont {Altland}\ and\ \citenamefont
  {Haake}(2012)}]{Altland2012}%
  \BibitemOpen
  \bibfield  {author} {\bibinfo {author} {\bibfnamefont {A.}~\bibnamefont
  {Altland}}\ and\ \bibinfo {author} {\bibfnamefont {F.}~\bibnamefont
  {Haake}},\ }\href {\doibase 10.1103/PhysRevLett.108.073601} {\bibfield
  {journal} {\bibinfo  {journal} {Phys. Rev. Lett.}\ }\textbf {\bibinfo
  {volume} {108}},\ \bibinfo {pages} {073601} (\bibinfo {year}
  {2012})}\BibitemShut {NoStop}%
\bibitem [{\citenamefont {Cucchietti}\ \emph {et~al.}(2002)\citenamefont
  {Cucchietti}, \citenamefont {Lewenkopf}, \citenamefont {Mucciolo},
  \citenamefont {Pastawski},\ and\ \citenamefont {Vallejos}}]{Cucchietti2002}%
  \BibitemOpen
  \bibfield  {author} {\bibinfo {author} {\bibfnamefont {F.~M.}\ \bibnamefont
  {Cucchietti}}, \bibinfo {author} {\bibfnamefont {C.~H.}\ \bibnamefont
  {Lewenkopf}}, \bibinfo {author} {\bibfnamefont {E.~R.}\ \bibnamefont
  {Mucciolo}}, \bibinfo {author} {\bibfnamefont {H.~M.}\ \bibnamefont
  {Pastawski}}, \ and\ \bibinfo {author} {\bibfnamefont {R.~O.}\ \bibnamefont
  {Vallejos}},\ }\href {\doibase 10.1103/PhysRevE.65.046209} {\bibfield
  {journal} {\bibinfo  {journal} {Phys. Rev. E}\ }\textbf {\bibinfo {volume}
  {65}},\ \bibinfo {pages} {046209} (\bibinfo {year} {2002})}\BibitemShut
  {NoStop}%
\bibitem [{\citenamefont {Lewis-Swan}\ \emph {et~al.}(2016)\citenamefont
  {Lewis-Swan}, \citenamefont {Olsen},\ and\ \citenamefont
  {Kheruntsyan}}]{Lewis-Swan2016}%
  \BibitemOpen
  \bibfield  {author} {\bibinfo {author} {\bibfnamefont {R.~J.}\ \bibnamefont
  {Lewis-Swan}}, \bibinfo {author} {\bibfnamefont {M.~K.}\ \bibnamefont
  {Olsen}}, \ and\ \bibinfo {author} {\bibfnamefont {K.~V.}\ \bibnamefont
  {Kheruntsyan}},\ }\href {\doibase 10.1103/PhysRevA.94.033814} {\bibfield
  {journal} {\bibinfo  {journal} {Phys. Rev. A}\ }\textbf {\bibinfo {volume}
  {94}},\ \bibinfo {pages} {033814} (\bibinfo {year} {2016})}\BibitemShut
  {NoStop}%
\bibitem [{\citenamefont {Bhattacharyya}(1943)}]{Bhatta1943}%
  \BibitemOpen
  \bibfield  {author} {\bibinfo {author} {\bibfnamefont {A.}~\bibnamefont
  {Bhattacharyya}},\ }\href@noop {} {\bibfield  {journal} {\bibinfo  {journal}
  {Bull. Calcutta Math. Soc.}\ }\textbf {\bibinfo {volume} {35}},\ \bibinfo
  {pages} {99} (\bibinfo {year} {1943})}\BibitemShut {NoStop}%
\bibitem [{\citenamefont {Tomkovi\v{c}}\ \emph {et~al.}(2017)\citenamefont
  {Tomkovi\v{c}}, \citenamefont {Muessel}, \citenamefont {Strobel},
  \citenamefont {L\"ock}, \citenamefont {Schlagheck}, \citenamefont
  {Ketzmerick},\ and\ \citenamefont {Oberthaler}}]{Tomkovic2017}%
  \BibitemOpen
  \bibfield  {author} {\bibinfo {author} {\bibfnamefont {J.}~\bibnamefont
  {Tomkovi\v{c}}}, \bibinfo {author} {\bibfnamefont {W.}~\bibnamefont
  {Muessel}}, \bibinfo {author} {\bibfnamefont {H.}~\bibnamefont {Strobel}},
  \bibinfo {author} {\bibfnamefont {S.}~\bibnamefont {L\"ock}}, \bibinfo
  {author} {\bibfnamefont {P.}~\bibnamefont {Schlagheck}}, \bibinfo {author}
  {\bibfnamefont {R.}~\bibnamefont {Ketzmerick}}, \ and\ \bibinfo {author}
  {\bibfnamefont {M.~K.}\ \bibnamefont {Oberthaler}},\ }\href {\doibase
  10.1103/PhysRevA.95.011602} {\bibfield  {journal} {\bibinfo  {journal} {Phys.
  Rev. A}\ }\textbf {\bibinfo {volume} {95}},\ \bibinfo {pages} {011602}
  (\bibinfo {year} {2017})}\BibitemShut {NoStop}%
\bibitem [{\citenamefont {Pokharel}\ \emph {et~al.}(2018)\citenamefont
  {Pokharel}, \citenamefont {Misplon}, \citenamefont {Lynn}, \citenamefont
  {Duggins}, \citenamefont {Hallman}, \citenamefont {Anderson}, \citenamefont
  {Kapulkin},\ and\ \citenamefont {Pattanayak}}]{Pokharel2018}%
  \BibitemOpen
  \bibfield  {author} {\bibinfo {author} {\bibfnamefont {B.}~\bibnamefont
  {Pokharel}}, \bibinfo {author} {\bibfnamefont {M.~Z.~R.}\ \bibnamefont
  {Misplon}}, \bibinfo {author} {\bibfnamefont {W.}~\bibnamefont {Lynn}},
  \bibinfo {author} {\bibfnamefont {P.}~\bibnamefont {Duggins}}, \bibinfo
  {author} {\bibfnamefont {K.}~\bibnamefont {Hallman}}, \bibinfo {author}
  {\bibfnamefont {D.}~\bibnamefont {Anderson}}, \bibinfo {author}
  {\bibfnamefont {A.}~\bibnamefont {Kapulkin}}, \ and\ \bibinfo {author}
  {\bibfnamefont {A.~K.}\ \bibnamefont {Pattanayak}},\ }\href {\doibase
  10.1038/s41598-018-20507-w} {\bibfield  {journal} {\bibinfo  {journal} {Sci.
  Rep.}\ }\textbf {\bibinfo {volume} {8}},\ \bibinfo {pages} {2108} (\bibinfo
  {year} {2018})}\BibitemShut {NoStop}%
\bibitem [{\citenamefont {Eastman}\ \emph {et~al.}(2019)\citenamefont
  {Eastman}, \citenamefont {Szigeti}, \citenamefont {Hope},\ and\ \citenamefont
  {Carvalho}}]{Eastman2019}%
  \BibitemOpen
  \bibfield  {author} {\bibinfo {author} {\bibfnamefont {J.~K.}\ \bibnamefont
  {Eastman}}, \bibinfo {author} {\bibfnamefont {S.~S.}\ \bibnamefont
  {Szigeti}}, \bibinfo {author} {\bibfnamefont {J.~J.}\ \bibnamefont {Hope}}, \
  and\ \bibinfo {author} {\bibfnamefont {A.~R.~R.}\ \bibnamefont {Carvalho}},\
  }\href {\doibase 10.1103/PhysRevA.99.012111} {\bibfield  {journal} {\bibinfo
  {journal} {Phys. Rev. A}\ }\textbf {\bibinfo {volume} {99}},\ \bibinfo
  {pages} {012111} (\bibinfo {year} {2019})}\BibitemShut {NoStop}%
\bibitem [{\citenamefont {Mahmud}\ \emph {et~al.}(2005)\citenamefont {Mahmud},
  \citenamefont {Perry},\ and\ \citenamefont {Reinhardt}}]{Mahmud2005}%
  \BibitemOpen
  \bibfield  {author} {\bibinfo {author} {\bibfnamefont {K.~W.}\ \bibnamefont
  {Mahmud}}, \bibinfo {author} {\bibfnamefont {H.}~\bibnamefont {Perry}}, \
  and\ \bibinfo {author} {\bibfnamefont {W.~P.}\ \bibnamefont {Reinhardt}},\
  }\href {\doibase 10.1103/PhysRevA.71.023615} {\bibfield  {journal} {\bibinfo
  {journal} {Phys. Rev. A}\ }\textbf {\bibinfo {volume} {71}},\ \bibinfo
  {pages} {023615} (\bibinfo {year} {2005})}\BibitemShut {NoStop}%
\bibitem [{\citenamefont {Trimborn}\ \emph {et~al.}(2009)\citenamefont
  {Trimborn}, \citenamefont {Witthaut},\ and\ \citenamefont
  {Korsch}}]{Trimborn2009}%
  \BibitemOpen
  \bibfield  {author} {\bibinfo {author} {\bibfnamefont {F.}~\bibnamefont
  {Trimborn}}, \bibinfo {author} {\bibfnamefont {D.}~\bibnamefont {Witthaut}},
  \ and\ \bibinfo {author} {\bibfnamefont {H.~J.}\ \bibnamefont {Korsch}},\
  }\href {\doibase 10.1103/PhysRevA.79.013608} {\bibfield  {journal} {\bibinfo
  {journal} {Phys. Rev. A}\ }\textbf {\bibinfo {volume} {79}},\ \bibinfo
  {pages} {013608} (\bibinfo {year} {2009})}\BibitemShut {NoStop}%
\bibitem [{\citenamefont {Hennig}\ \emph {et~al.}(2012)\citenamefont {Hennig},
  \citenamefont {Witthaut},\ and\ \citenamefont {Campbell}}]{Hennig2012}%
  \BibitemOpen
  \bibfield  {author} {\bibinfo {author} {\bibfnamefont {H.}~\bibnamefont
  {Hennig}}, \bibinfo {author} {\bibfnamefont {D.}~\bibnamefont {Witthaut}}, \
  and\ \bibinfo {author} {\bibfnamefont {D.~K.}\ \bibnamefont {Campbell}},\
  }\href {\doibase 10.1103/PhysRevA.86.051604} {\bibfield  {journal} {\bibinfo
  {journal} {Phys. Rev. A}\ }\textbf {\bibinfo {volume} {86}},\ \bibinfo
  {pages} {051604} (\bibinfo {year} {2012})}\BibitemShut {NoStop}%
\bibitem [{\citenamefont {Milburn}\ \emph
  {et~al.}(1997{\natexlab{a}})\citenamefont {Milburn}, \citenamefont {Corney},
  \citenamefont {Wright},\ and\ \citenamefont {Walls}}]{Milburn1997}%
  \BibitemOpen
  \bibfield  {author} {\bibinfo {author} {\bibfnamefont {G.~J.}\ \bibnamefont
  {Milburn}}, \bibinfo {author} {\bibfnamefont {J.}~\bibnamefont {Corney}},
  \bibinfo {author} {\bibfnamefont {E.}~\bibnamefont {Wright}}, \ and\ \bibinfo
  {author} {\bibfnamefont {D.~F.}\ \bibnamefont {Walls}},\ }\href {\doibase
  10.1103/PhysRevA.55.4318} {\bibfield  {journal} {\bibinfo  {journal} {Phys.
  Rev. A}\ }\textbf {\bibinfo {volume} {55}},\ \bibinfo {pages} {4318}
  (\bibinfo {year} {1997}{\natexlab{a}})}\BibitemShut {NoStop}%
\bibitem [{\citenamefont {Milburn}\ \emph
  {et~al.}(1997{\natexlab{b}})\citenamefont {Milburn}, \citenamefont {Corney},
  \citenamefont {Harris}, \citenamefont {Wright},\ and\ \citenamefont
  {Walls}}]{Milburn1997chaos}%
  \BibitemOpen
  \bibfield  {author} {\bibinfo {author} {\bibfnamefont {G.~J.}\ \bibnamefont
  {Milburn}}, \bibinfo {author} {\bibfnamefont {J.~F.}\ \bibnamefont {Corney}},
  \bibinfo {author} {\bibfnamefont {D.}~\bibnamefont {Harris}}, \bibinfo
  {author} {\bibfnamefont {E.~M.}\ \bibnamefont {Wright}}, \ and\ \bibinfo
  {author} {\bibfnamefont {D.~F.}\ \bibnamefont {Walls}},\ }in\ \href {\doibase
  10.1117/12.273762} {\emph {\bibinfo {booktitle} {Proc. SPIE}}},\ Vol.\
  \bibinfo {volume} {2995},\ \bibinfo {editor} {edited by\ \bibinfo {editor}
  {\bibfnamefont {M.~G.}\ \bibnamefont {Prentiss}}\ and\ \bibinfo {editor}
  {\bibfnamefont {W.~D.}\ \bibnamefont {Phillips}}}\ (\bibinfo  {publisher}
  {SPIE},\ \bibinfo {year} {1997})\ pp.\ \bibinfo {pages}
  {232--239}\BibitemShut {NoStop}%
\bibitem [{\citenamefont {Albiez}\ \emph {et~al.}(2005)\citenamefont {Albiez},
  \citenamefont {Gati}, \citenamefont {F\"olling}, \citenamefont {Hunsmann},
  \citenamefont {Cristiani},\ and\ \citenamefont {Oberthaler}}]{Albiez2005}%
  \BibitemOpen
  \bibfield  {author} {\bibinfo {author} {\bibfnamefont {M.}~\bibnamefont
  {Albiez}}, \bibinfo {author} {\bibfnamefont {R.}~\bibnamefont {Gati}},
  \bibinfo {author} {\bibfnamefont {J.}~\bibnamefont {F\"olling}}, \bibinfo
  {author} {\bibfnamefont {S.}~\bibnamefont {Hunsmann}}, \bibinfo {author}
  {\bibfnamefont {M.}~\bibnamefont {Cristiani}}, \ and\ \bibinfo {author}
  {\bibfnamefont {M.~K.}\ \bibnamefont {Oberthaler}},\ }\href {\doibase
  10.1103/PhysRevLett.95.010402} {\bibfield  {journal} {\bibinfo  {journal}
  {Phys. Rev. Lett.}\ }\textbf {\bibinfo {volume} {95}},\ \bibinfo {pages}
  {010402} (\bibinfo {year} {2005})}\BibitemShut {NoStop}%
\bibitem [{\citenamefont {Mourik}\ \emph {et~al.}(2018)\citenamefont {Mourik},
  \citenamefont {Asaad}, \citenamefont {Firgau}, \citenamefont {Pla},
  \citenamefont {Holmes}, \citenamefont {Milburn}, \citenamefont {McCallum},\
  and\ \citenamefont {Morello}}]{Mourik2018}%
  \BibitemOpen
  \bibfield  {author} {\bibinfo {author} {\bibfnamefont {V.}~\bibnamefont
  {Mourik}}, \bibinfo {author} {\bibfnamefont {S.}~\bibnamefont {Asaad}},
  \bibinfo {author} {\bibfnamefont {H.}~\bibnamefont {Firgau}}, \bibinfo
  {author} {\bibfnamefont {J.~J.}\ \bibnamefont {Pla}}, \bibinfo {author}
  {\bibfnamefont {C.}~\bibnamefont {Holmes}}, \bibinfo {author} {\bibfnamefont
  {G.~J.}\ \bibnamefont {Milburn}}, \bibinfo {author} {\bibfnamefont {J.~C.}\
  \bibnamefont {McCallum}}, \ and\ \bibinfo {author} {\bibfnamefont
  {A.}~\bibnamefont {Morello}},\ }\href {\doibase 10.1103/PhysRevE.98.042206}
  {\bibfield  {journal} {\bibinfo  {journal} {Phys. Rev. E}\ }\textbf {\bibinfo
  {volume} {98}},\ \bibinfo {pages} {042206} (\bibinfo {year}
  {2018})}\BibitemShut {NoStop}%
\bibitem [{\citenamefont {Gong}\ \emph {et~al.}(2009)\citenamefont {Gong},
  \citenamefont {Morales-Molina},\ and\ \citenamefont
  {H{\"{a}}nggi}}]{Gong2009}%
  \BibitemOpen
  \bibfield  {author} {\bibinfo {author} {\bibfnamefont {J.}~\bibnamefont
  {Gong}}, \bibinfo {author} {\bibfnamefont {L.}~\bibnamefont
  {Morales-Molina}}, \ and\ \bibinfo {author} {\bibfnamefont {P.}~\bibnamefont
  {H{\"{a}}nggi}},\ }\href {\doibase 10.1103/PhysRevLett.103.133002} {\bibfield
   {journal} {\bibinfo  {journal} {Phys. Rev. Lett.}\ }\textbf {\bibinfo
  {volume} {103}},\ \bibinfo {pages} {133002} (\bibinfo {year}
  {2009})}\BibitemShut {NoStop}%
\bibitem [{\citenamefont {Watanabe}\ and\ \citenamefont
  {M\"akel\"a}(2012)}]{Watanabe2012}%
  \BibitemOpen
  \bibfield  {author} {\bibinfo {author} {\bibfnamefont {G.}~\bibnamefont
  {Watanabe}}\ and\ \bibinfo {author} {\bibfnamefont {H.}~\bibnamefont
  {M\"akel\"a}},\ }\href {\doibase 10.1103/PhysRevA.85.053624} {\bibfield
  {journal} {\bibinfo  {journal} {Phys. Rev. A}\ }\textbf {\bibinfo {volume}
  {85}},\ \bibinfo {pages} {053624} (\bibinfo {year} {2012})}\BibitemShut
  {NoStop}%
\bibitem [{\citenamefont {Morales-Molina}\ and\ \citenamefont
  {Gong}(2008)}]{Morales-Molina2008}%
  \BibitemOpen
  \bibfield  {author} {\bibinfo {author} {\bibfnamefont {L.}~\bibnamefont
  {Morales-Molina}}\ and\ \bibinfo {author} {\bibfnamefont {J.}~\bibnamefont
  {Gong}},\ }\href {\doibase 10.1103/PhysRevA.78.041403} {\bibfield  {journal}
  {\bibinfo  {journal} {Phys. Rev. A}\ }\textbf {\bibinfo {volume} {78}},\
  \bibinfo {pages} {1} (\bibinfo {year} {2008})}\BibitemShut {NoStop}%
\bibitem [{\citenamefont {Chianca}\ and\ \citenamefont
  {Olsen}(2011)}]{Chianca2011}%
  \BibitemOpen
  \bibfield  {author} {\bibinfo {author} {\bibfnamefont {C.~V.}\ \bibnamefont
  {Chianca}}\ and\ \bibinfo {author} {\bibfnamefont {M.~K.}\ \bibnamefont
  {Olsen}},\ }\href {\doibase 10.1103/PhysRevA.84.043636} {\bibfield  {journal}
  {\bibinfo  {journal} {Phys. Rev. A}\ }\textbf {\bibinfo {volume} {84}},\
  \bibinfo {pages} {043636} (\bibinfo {year} {2011})}\BibitemShut {NoStop}%
\bibitem [{\citenamefont {Kohler}\ \emph {et~al.}(2005)\citenamefont {Kohler},
  \citenamefont {Lehmann},\ and\ \citenamefont {H{\"{a}}nggi}}]{Kohler2005}%
  \BibitemOpen
  \bibfield  {author} {\bibinfo {author} {\bibfnamefont {S.}~\bibnamefont
  {Kohler}}, \bibinfo {author} {\bibfnamefont {J.}~\bibnamefont {Lehmann}}, \
  and\ \bibinfo {author} {\bibfnamefont {P.}~\bibnamefont {H{\"{a}}nggi}},\
  }\href {\doibase 10.1016/j.physrep.2004.11.002} {\bibfield  {journal}
  {\bibinfo  {journal} {Phys. Rep.}\ }\textbf {\bibinfo {volume} {406}},\
  \bibinfo {pages} {379} (\bibinfo {year} {2005})}\BibitemShut {NoStop}%
\bibitem [{\citenamefont {Salmond}\ \emph {et~al.}(2002)\citenamefont
  {Salmond}, \citenamefont {Holmes},\ and\ \citenamefont
  {Milburn}}]{Salmond2002}%
  \BibitemOpen
  \bibfield  {author} {\bibinfo {author} {\bibfnamefont {G.~L.}\ \bibnamefont
  {Salmond}}, \bibinfo {author} {\bibfnamefont {C.~A.}\ \bibnamefont {Holmes}},
  \ and\ \bibinfo {author} {\bibfnamefont {G.~J.}\ \bibnamefont {Milburn}},\
  }\href {\doibase 10.1103/PhysRevA.65.033623} {\bibfield  {journal} {\bibinfo
  {journal} {Phys. Rev. A}\ }\textbf {\bibinfo {volume} {65}},\ \bibinfo
  {pages} {033623} (\bibinfo {year} {2002})}\BibitemShut {NoStop}%
\bibitem [{\citenamefont {Arecchi}\ \emph {et~al.}(1972)\citenamefont
  {Arecchi}, \citenamefont {Courtens}, \citenamefont {Gilmore},\ and\
  \citenamefont {Thomas}}]{Arecchi1972}%
  \BibitemOpen
  \bibfield  {author} {\bibinfo {author} {\bibfnamefont {F.~T.}\ \bibnamefont
  {Arecchi}}, \bibinfo {author} {\bibfnamefont {E.}~\bibnamefont {Courtens}},
  \bibinfo {author} {\bibfnamefont {R.}~\bibnamefont {Gilmore}}, \ and\
  \bibinfo {author} {\bibfnamefont {H.}~\bibnamefont {Thomas}},\ }\href
  {\doibase 10.1103/PhysRevA.6.2211} {\bibfield  {journal} {\bibinfo  {journal}
  {Phys. Rev. A}\ }\textbf {\bibinfo {volume} {6}},\ \bibinfo {pages} {2211}
  (\bibinfo {year} {1972})}\BibitemShut {NoStop}%
\bibitem [{\citenamefont {Lee}\ and\ \citenamefont {Feit}(1993)}]{Lee1993}%
  \BibitemOpen
  \bibfield  {author} {\bibinfo {author} {\bibfnamefont {S.~B.}\ \bibnamefont
  {Lee}}\ and\ \bibinfo {author} {\bibfnamefont {M.~D.}\ \bibnamefont {Feit}},\
  }\href {\doibase 10.1103/PhysRevE.47.4552} {\bibfield  {journal} {\bibinfo
  {journal} {Phys. Rev. E}\ }\textbf {\bibinfo {volume} {47}},\ \bibinfo
  {pages} {4552} (\bibinfo {year} {1993})}\BibitemShut {NoStop}%
\bibitem [{Note1()}]{Note1}%
  \BibitemOpen
  \bibinfo {note} {Bloch coherent states~\cite {Arecchi1972} are minimal
  uncertainty states of the dimer under the restriction of fixed total particle
  number. They may be created experimentally by, for example, forming a
  condensate of known number in one well, followed by linear tunnelling~\cite
  {Tomkovic2017}.}\BibitemShut {Stop}%
\bibitem [{\citenamefont {Leggett}(2001)}]{Leggett2001}%
  \BibitemOpen
  \bibfield  {author} {\bibinfo {author} {\bibfnamefont {A.~J.}\ \bibnamefont
  {Leggett}},\ }\href {\doibase 10.1103/RevModPhys.73.307} {\bibfield
  {journal} {\bibinfo  {journal} {Rev. Mod. Phys.}\ }\textbf {\bibinfo {volume}
  {73}},\ \bibinfo {pages} {307} (\bibinfo {year} {2001})}\BibitemShut
  {NoStop}%
\bibitem [{\citenamefont {Graham}(1973)}]{Graham1973}%
  \BibitemOpen
  \bibfield  {author} {\bibinfo {author} {\bibfnamefont {R.}~\bibnamefont
  {Graham}},\ }\href {\doibase 10.1007/978-3-662-40468-3} {\bibfield  {journal}
  {\bibinfo  {journal} {Springer Tr. Mod. Phys.}\ }\textbf {\bibinfo {volume}
  {66}},\ \bibinfo {pages} {1} (\bibinfo {year} {1973})}\BibitemShut {NoStop}%
\bibitem [{\citenamefont {Gardiner}\ and\ \citenamefont
  {Collett}(1985)}]{Gardiner1985}%
  \BibitemOpen
  \bibfield  {author} {\bibinfo {author} {\bibfnamefont {C.~W.}\ \bibnamefont
  {Gardiner}}\ and\ \bibinfo {author} {\bibfnamefont {M.~J.}\ \bibnamefont
  {Collett}},\ }\href {\doibase 10.1103/PhysRevA.31.3761} {\bibfield  {journal}
  {\bibinfo  {journal} {Phys. Rev. A}\ }\textbf {\bibinfo {volume} {31}},\
  \bibinfo {pages} {3761} (\bibinfo {year} {1985})}\BibitemShut {NoStop}%
\bibitem [{\citenamefont {Olsen}\ and\ \citenamefont
  {Bradley}(2009)}]{Olsen2009}%
  \BibitemOpen
  \bibfield  {author} {\bibinfo {author} {\bibfnamefont {M.~K.}\ \bibnamefont
  {Olsen}}\ and\ \bibinfo {author} {\bibfnamefont {A.~S.}\ \bibnamefont
  {Bradley}},\ }\href {\doibase 10.1016/j.optcom.2009.06.033} {\bibfield
  {journal} {\bibinfo  {journal} {Opt. Commun.}\ }\textbf {\bibinfo {volume}
  {282}},\ \bibinfo {pages} {3924} (\bibinfo {year} {2009})}\BibitemShut
  {NoStop}%
\bibitem [{\citenamefont {Corney}\ and\ \citenamefont
  {Olsen}(2015)}]{Corney2015}%
  \BibitemOpen
  \bibfield  {author} {\bibinfo {author} {\bibfnamefont {J.~F.}\ \bibnamefont
  {Corney}}\ and\ \bibinfo {author} {\bibfnamefont {M.~K.}\ \bibnamefont
  {Olsen}},\ }\href {\doibase 10.1103/PhysRevA.91.023824} {\bibfield  {journal}
  {\bibinfo  {journal} {Phys. Rev. A}\ }\textbf {\bibinfo {volume} {91}},\
  \bibinfo {pages} {023824} (\bibinfo {year} {2015})}\BibitemShut {NoStop}%
\bibitem [{\citenamefont {Dennis}\ \emph {et~al.}(2013)\citenamefont {Dennis},
  \citenamefont {Hope},\ and\ \citenamefont {Johnsson}}]{Dennis2013}%
  \BibitemOpen
  \bibfield  {author} {\bibinfo {author} {\bibfnamefont {G.~R.}\ \bibnamefont
  {Dennis}}, \bibinfo {author} {\bibfnamefont {J.~J.}\ \bibnamefont {Hope}}, \
  and\ \bibinfo {author} {\bibfnamefont {M.~T.}\ \bibnamefont {Johnsson}},\
  }\href {\doibase http://dx.doi.org/10.1016/j.cpc.2012.08.016} {\bibfield
  {journal} {\bibinfo  {journal} {Comput. Phys. Commun.}\ }\textbf {\bibinfo
  {volume} {184}},\ \bibinfo {pages} {201} (\bibinfo {year}
  {2013})}\BibitemShut {NoStop}%
\bibitem [{\citenamefont {Gra{\ss}}\ \emph {et~al.}(2013)\citenamefont
  {Gra{\ss}}, \citenamefont {Juli{\'{a}}-D{\'{i}}az}, \citenamefont {Ku{\aa}},\
  and\ \citenamefont {Lewenstein}}]{Grass2013}%
  \BibitemOpen
  \bibfield  {author} {\bibinfo {author} {\bibfnamefont {T.}~\bibnamefont
  {Gra{\ss}}}, \bibinfo {author} {\bibfnamefont {B.}~\bibnamefont
  {Juli{\'{a}}-D{\'{i}}az}}, \bibinfo {author} {\bibfnamefont {M.}~\bibnamefont
  {Ku{\aa}}}, \ and\ \bibinfo {author} {\bibfnamefont {M.}~\bibnamefont
  {Lewenstein}},\ }\href {\doibase 10.1103/PhysRevLett.111.090404} {\bibfield
  {journal} {\bibinfo  {journal} {Phys. Rev. Lett.}\ }\textbf {\bibinfo
  {volume} {111}},\ \bibinfo {pages} {1} (\bibinfo {year} {2013})}\BibitemShut
  {NoStop}%
\bibitem [{\citenamefont {Sakurai}\ and\ \citenamefont
  {Napolitano}(2014)}]{Sakurai2014}%
  \BibitemOpen
  \bibfield  {author} {\bibinfo {author} {\bibfnamefont {J.~J.}\ \bibnamefont
  {Sakurai}}\ and\ \bibinfo {author} {\bibfnamefont {J.~J.}\ \bibnamefont
  {Napolitano}},\ }\href@noop {} {\emph {\bibinfo {title} {Modern Quantum
  Mechanics}}},\ \bibinfo {edition} {2nd}\ ed.\ (\bibinfo  {publisher} {Pearson
  Education Limited},\ \bibinfo {year} {2014})\BibitemShut {NoStop}%
\end{thebibliography}
\end{document}